\documentclass[a4paper,10pt, twocolumn]{article}

\usepackage[top=1.5cm, left=1.5cm, right=1.5cm, bottom=1.5cm]{geometry}

\usepackage{amsfonts,amssymb,amsmath,amsthm,dsfont}

\usepackage{subfigure}
\usepackage{color}

\usepackage{graphicx}

\makeatletter
\let\NAT@parse\undefined
\makeatother
\usepackage[sort&compress, numbers]{natbib}

\usepackage{algorithm,algorithmic}
\usepackage[ruled,vlined,algosection,algo2e]{algorithm2e}  
\graphicspath{{./},{images/}}
\usepackage{hyperref}

\usepackage[footnotesize]{caption}

\newcommand{\bs}{\boldsymbol}

\newcommand{\ie}{\emph{i.e.}, }

\usepackage{comment}

\makeatletter
\renewcommand{\section}{\@startsection {section}{1}{\z@}%
{-3.5ex \@plus -1ex \@minus -.2ex}%
{2.3ex \@plus.2ex}%
{\normalfont\large\bfseries}}
\makeatother 

\title{ \vspace{-0.5cm}Reference-less algorithm for circumstellar disks imaging}
\author{Beno\^{i}t Pairet$^1$\thanks{BP and LJ are funded by the Belgian F.R.S.-FNRS. Part of this study is funded by the project AlterSense (MIS-FNRS)}, Faustine Cantalloube$^2$, Laurent Jacques$^{1*}$\\
\footnotesize $^1$ISPGroup, ICTEAM/ELEN, UCLouvain, Belgium\ $^2$ Max Planck Institute for Astronomy, Germany
}

\date{\empty}

\renewenvironment{abstract}{\noindent\bf\small {\em Abstract---}}{}

\begin{document}

\maketitle

\begin{abstract}
Circumstellar disks play a key role in the understanding of stellar systems. Direct imaging of such extended structures is a challenging task. Current post-processing techniques, first tailored for exoplanets imaging, tend to produce deformed images of the circumstellar disks, hindering our capability to study their shape and photometry in details. We address here the reasons of this shortcoming and propose an algorithm that produces more faithful images of disks taken with ground-based telescopes. We also show that our algorithm is a good candidate for exoplanets imaging. We then explain how our approach can be extended in the form of a regularized inverse problem.
\end{abstract}

\section{Introduction}
 Imaging stellar systems requires high contrast and high resolution to resolve the faint objects close to the much brighter hosting star. Ground based telescopes offer the highest resolution but suffer from atmospheric turbulence. Adaptive optics is used to correct the wavefront, thus allowing the use of a coronagraph which hides most of the starlight to unveil faint surrounding structures. Even with state-of-the-art hardware, there are remaining quasi-statics aberrations forming  speckles in the image, which prevent detection of faint signals. Specific observation methods are used to increase data diversity to disentangle the on-sky signals from the speckles.
Angular differential imaging (ADI) is a popular observation method~\cite{marois2006angular} where multiple snapshots of the star are taken through a night of observation in a way such that the speckles remain quasi-static while on-sky signals follow a deterministic circular trajectory, determined by the parallactic angles. Dedicated processing is then required to process ADI datasets. Among them, Principal components analysis (PCA) is a popular method~\cite{soummer2012detection,amara2012pynpoint}. The reduction procedure, \ie the processing pipeline, goes as follows, \textit{(i)} the $n\times n \times t$ spatiotemporal data cube is reshaped into a $\mathbb R^{T\times n^2}$ matrix $\bs Y$, \textit{(ii)}  we compute its low rank approximation, $\bs L$ through an SVD, \textit{(iii)} $\bs L$ is subtracted from the data to form $\bs S = \bs Y - \bs L$ containing the on-sky signals, and \textit{(iv)} the frames of $\bs S$ are aligned to a common direction for the on-sky signal and temporally averaged to form the processed frame, we denote this last step as $\text{Red}(\bs S)$. A detection is then usually performed either directly on the processed frame~\cite{mawet2014fundamental} or using the volume $\bs S$~\cite{pairet2018stim}. Throughout this paper, we write the best $r$ rank approximation of a matrix $\bs X$ as $\mathcal H^{\text{SVD}}_r(\bs X)$. 

Although ADI-based methods proved to be a powerful tool for exoplanets detection, observing circumstellar disks remains an uneasy task as the shape of the disks is not known and can be quite irregular with smooth edges. The morphology of the disks is known to be severely impacted by the reduction procedure, hindering our capability to study the structure of the disks from ADI datasets. Very few attempts of tackling the problem of the disk deformation induced by the processing on ADI datasets have been made so far. We mention~\cite{milli2012impact}, in which the authors used a forward modeling of disks that they inject on the dataset to estimate the deformation induced by PCA. Another promising approach is found in~\cite{ren2018non} where the authors impose positivity on the processed frame by means of non-negative matrix factorization. 

\section{Our approach}
\begin{figure}[t]
  \centering
  {\includegraphics[width=0.21\textwidth]{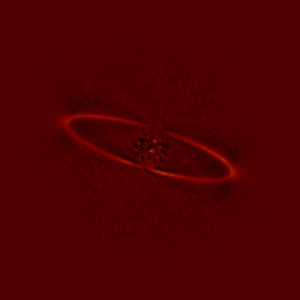}}
 \hspace{1mm} 
   {\includegraphics[width=0.21\textwidth]{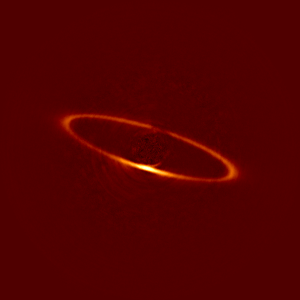}}\\
      \vspace{2mm} 
  {\includegraphics[width=0.21\textwidth]{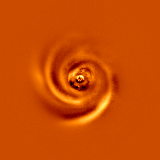}}
   \hspace{1mm} 
  {\includegraphics[width=0.21\textwidth]{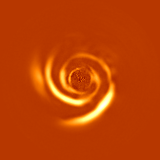}}\\
      \vspace{2mm} 
    {\includegraphics[width=0.21\textwidth]{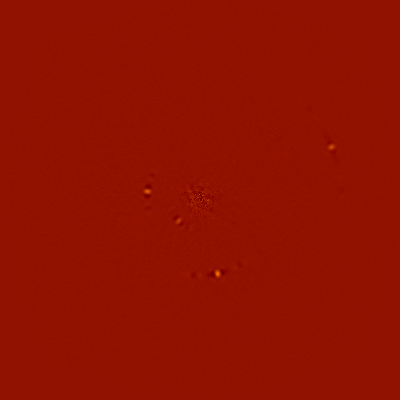}}
 \hspace{2mm} 
  {\includegraphics[width=0.21\textwidth]{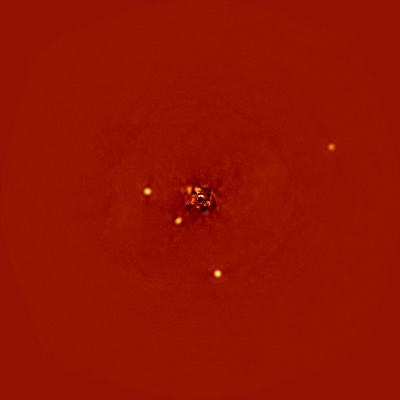}}
\caption{Processed frames using PCA (left) and our algorithm (right) for HR~4796A (top), SAO~206462 (middle), and HR~8799 (bottom). The colorbars are arbitrary but are consistent between left and right. Both disks are deformed in the PCA-processing. The shapes of the disks are closer to their actual shapes, obtained using polarimetric differential imaging. For HR~8799, the processed frame exhibits the typical PCA-induced deformation: two negative negative regions next to the planets. On the other hand, our algorithm yields a frame where planets have a round shape and we can even see the second ring of the Airy disk. }
\label{fig:pca_vs_our_algorithm}
\vskip -3ex
\end{figure}

In this short paper, we investigate the reasons behind the failure of PCA and provide an algorithm to overcome this shortcoming. First we illustrate the deformation induced by PCA on bright disks taken with the SPHERE high-contrast instrument~\cite{beuzit2008sphere} installed at the Very Large Telescope in Chile. The first is the ellipsoidal disk surrounding HR~4796A~\cite{milli2017near}. The second is the sprial disk surrounding SAO~206462~\cite{maire2017testing}.
Fig.~\ref{fig:pca_vs_our_algorithm} shows the images of these disks obtained with PCA (top and middle, left). The disk around HR~4796A is supposed to have an ellipsoidal shape, but here we see negative (in dark) rotated copies of the disk on both sides of it. For SAO~206462, polarized images show its actual form (see~\cite{garufi2013small}), we also see negative artifacts surrounding the signal of the disk which is distorted by the PCA process.
The processing relies on the SVD where $\bs Y = \bs U \bs \Sigma \bs V^\top$. In our setting, the matrix $\bs V \in \mathbb R^{T \times n^2}$ is a list of $T$ $n \times n$ images (each images being reshaped into a vector in $\mathbb R^{n^2 }$, by a slight abuse of notation, we call them images throughout this paper), $\bs \Sigma$ is a list of weight attributed to each image, and the time variation is encoded in $\bs U$ where different weights are assigned to each image of $\bs V$ through the temporal evolution. 

By analyzing the images encoded in $\bs V$, we see that the way the disk is encoded in these images causes the deformation in the PCA processed frame. This is emphasized when building $\bs S$ as the polar decomposition of $\bs Y$, \ie as $\bs S = \bs U \bs V^\top$, where we see the same kind of deformations without any ``self-absorption''. 

\begin{figure}
  \centering
  {\includegraphics[width=0.21\textwidth]{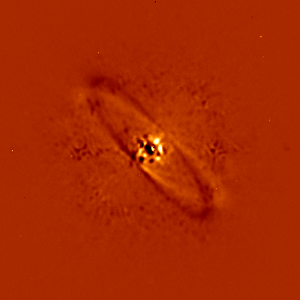}}
 \hspace{2mm} 
  {\includegraphics[width=0.21\textwidth]{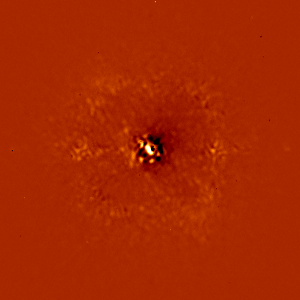}}
\caption{Difference in the morphology of the images of V. Left, sixth image of $\bs V$ built from $\bs Y$. Rigth, sixth image of $\bs V$ built from $\bs Y - \bs D$. Where $\bs D$ is the estimate of the disk. We can see that for the latter case, the intensity of the disk is significantly lower.}
\label{fig:V_init_vs_V_algo}
\vskip -3ex
\end{figure}

Because the disk rotates, it has to be present in the images sequence at different positions. Unfortunately, the reference images $\bs V$ do not encode the temporal evolution. Hence the disk must be represented in $\bs V$ with positive and negative rotated copies whose linear combinations explain the temporal evolution of the disk. The left image of Fig.~\ref{fig:V_init_vs_V_algo} shows the 6$^{\text{th}}$ image of $\bs V$, where we can see this effect clearly.

The proposed algorithm iteratively subtracts the best estimate of the disk from the data so its impact onto the images of $\bs V$ decreases.  
To motivate this, we show that using an estimate from a PCA-processed frame significantly reduces the impact of the disk on $\bs V$. We compute the processed frame with $\bs L = \mathcal H^{\text{SVD}}_1(\bs Y)$ and keep only the positive parts of it and denote it $\bs d$. We then inject it in a $t\times n \times n$ cube and rotate the frames according to the parallactic angles, we denote this operation by $\bs D = \Theta(\bs d)$. Fig.~\ref{fig:V_init_vs_V_algo} (right) displays an image of $\bs V$ obtained by the SVD of $\bs Y - \bs D$. We can see that in this case the disk structure does not appear as clearly as on Fig.~\ref{fig:V_init_vs_V_algo} (left).

Then the algorithm iterates with $r=1,\dots,p$ and we compute $\bs S^{(r)} =  \bs Y - \mathcal H_r^{\text{SVD}}(\bs Y - \Theta(\text{Red}(\bs S^{(r-1)}))$. This is motivated by the idea that if a better approximation of the disk is removed from $\bs Y$, its impact on $\bs V$ further decreases. In return if the disk has a lower impact on $\bs V$, we expect to get a better approximation of the disk. 

The rest of the algorithm follows the classical ADI reduction, $\bs S^{(p)}$ is aligned and collapsed to form the processed frame $\bs f$. The full algorithm is summarized in Alg.~\ref{alg:Algo_1}. Fig.~\ref{fig:pca_vs_our_algorithm} (top and middle, right) diplays the results of this algorithm applied to the two datasets shown earlier. For SAO~206462, comparing with results displayed on Figure 1 of~\cite{garufi2013small}, one can see that the proposed algorithm produced a well-rendered shape.

\begin{algorithm}[b!]
\small
\caption{\small }
\begin{algorithmic}[1]
 \STATE \textbf{Input:} $\bs Y$, $\bs m$, $p$\\
 \STATE \textbf{Output:} $\bs f$\\
  \STATE $\bs S^{(0)} =  \bs Y - \mathcal H_1^{\text{SVD}}(\bs Y)$\\
      \FOR{$r=1,2,\dots, p$}
  \STATE $S^{(r)} = \bs Y - \mathcal H_r^{\text{SVD}} (\bs Y - \Theta(\text{Red}(\bs S^{(r-1)})))$\\
      \ENDFOR
      \STATE $\textstyle  \bs f =\text{Red}(\bs S^{(p)})$
\end{algorithmic}
\label{alg:Algo_1}
\end{algorithm}

To further demonstrate the capability of our algorithm to preserve shapes, we applied it to exoplanets imaging, on the SPHERE data of HR~8799. For exoplanets, we know that the shape of the planets is similar to an Airy disk. We display the PCA-processed frame on~Fig.~\ref{fig:pca_vs_our_algorithm} (bottom left). We observe the typical PCA-induced shape, the planet is surrounded by two negative regions. The results of our approach is displayed on Fig.~\ref{fig:pca_vs_our_algorithm} (bottom right). By a carful inspection, we can see that the Airy-disk like shape is rendered by our method, hence indicating that it is indeed shape preserving.

\section{Conclusion}
Alg.~\ref{alg:Algo_1} shows promising results for both disks and exoplanets imaging. This algorithm is similar to a greedy version of a regularized inverse problem. It would then probably benefit from being formulated as a convex optimization program and solved exactly. To do so, we need to impose a low rank structure for $\bs L$, using the nuclear norm and ``rotating 1-rank'' structure for $\bs S$. This could be done using the Radon domain where an image rotation corresponds to a shift. This way, the temporal evolution of $\bs S$, \ie the rotation of the same frame, is encoded as a convolution with circular boundary conditions. We observed that the Radon transform preserves the low rank structure of $\bs L$. Hence the algorithm could be fully described in the Radon domain. This more rigorous approach is left as a future work. 

We also showed that Alg.~\ref{alg:Algo_1} can be used for exoplanets detection. In the processed frame the shape of the exoplanets is recovered. This provides the opportunity to further disentangle the exoplanet signal from residual noise by means of a deconvolution. This is left as future work.

\textbf{Acknowledgements: } The authors thank Julien Milli and Anne-Lise Maire for kindly providing the SPHERE datasets used in this study.

\clearpage

\small
\bibliography{biblio}{}
\bibliographystyle{plain}

\end{document}